# Performance of a MQXF Nb₃Sn Quadrupole Magnet Under Different Stress Level

Susana Izquierdo Bermudez, Giorgio Ambrosio, Bernardo Bordini, Nicolas Bourcey, Paolo Ferracin, Jose Ferradas Troitino, Salvador Ferradas Troitino, Lucio Fiscarelli, Jerome Fleiter, Michael Guinchard, Franco Mangiarotti, Juan Carlos Perez, Eelis Takala, Ezio Todesco.

*Abstract*— In a dipole or in a quadrupole accelerator magnet, the displacement of the coil turns induced by the electromagnetic forces can cause quenches limiting the magnet performance. For this reason, an azimuthal preload is applied to avoid azimuthal movements of the coil up to the required operational current. However, several tests showed that accelerator magnets can operate with a partial preload, i.e. that coil unloading during the ramp does not prevent reaching higher currents. This issue is particularly relevant for Nb₃Sn magnets, where the loads applied to the Nb₃Sn filaments can reach the degradation limits of critical current. In order to investigate the impact of coil preload on the quench performance, the MQXFS6 short model quadrupole for the High Luminosity Upgrade was tested under an azimuthal preload at 80% of the short sample current, reaching 93% of short sample current at 1.9 K. The preload was then released to 60%, still showing ability to operate in the range of 80-85% of short sample current as required by HL-LHC project. With this lower preload, the ability of going above 90% of short sample was lost, and a significant training appeared above 85%. When the preload was restored to the original 80% value, the magnet reached with few quenches 95% of short sample (13.4 T peak field). Magnetic measurements confirm the larger movement of the coil in the case with lower preload, and agree with finite element simulations.

## I. Introduction

IN a dipole or quadrupole accelerator magnet the electromagnetic forces in the coil are directed towards the mid plane and radially outwards (see Fig. 1). Displacement of the turns at powering can provoke variations of field quality during the ramp and cause releases of frictional energy, which could trigger a quench. The first effect was considered to be so critical that it triggered the preload strategies for the Tevatron main dipoles (see [1], mid of page 276). However, improvements in magnet powering and controls demonstrated that corrector magnets can compensate this effect and nowadays field quality is not a reason for preloading the coils. The second effect (quenches, training, and possibly performance limitation) is still valid, and the design criteria used in many projects is that the superconducting coil should be compressed ("preloaded") during assembly to avoid pole detachment during operational conditions. This principle was used in the LHC dipoles (see Ref. [2], end of page 169), and more recently for Nb₃Sn technology, as in the LARP HQ quadrupoles [3], in the HL-LHC 11 T dipole [4], in the HL-LHC MQXF quadrupole [5], and in the design of 16 T dipoles for the Future Circular Collider [6].

On the other hand, superconducting accelerator magnets have proven in many cases that coil unloading during the magnet ramp does not prevent reaching higher currents. This was observed for the SSC prototypes ([7], page 1358), in short models for the LHC dipoles [8], and more recently in short models for MQXF quadrupoles [9,10] and in short models for the 11 T dipole [10,11].

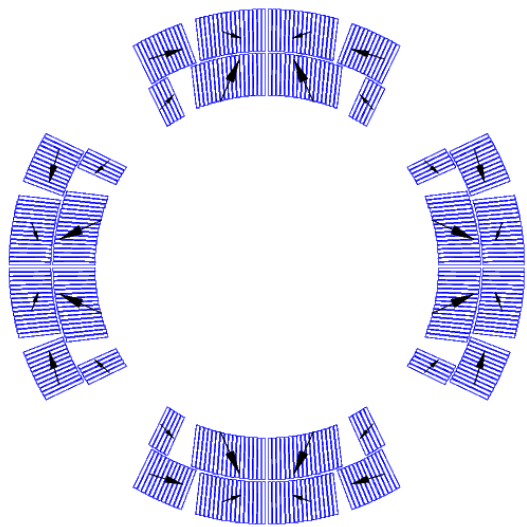

Fig. 1. MQXF coil lay-out including electromagnetic forces

The guidelines for coil preload are particularly relevant for high field or large aperture Nb₃Sn magnets, since stress scales with aperture, field and current density ([12], page 6), and the requirement on full preload risks to approach the degradation levels of the superconductor. In order to investigate the impact

Manuscript received XX;
Susana Izquierdo Bermudez ; Marta Bajko ; Bernardo Bordini ; Nicolas Bourcey ; Delio Duarte Ramos,; Lucio Fiscarelli Jerome Fleiter ; Michael Guinchard ; Friedrich Lackner ; Nicholas Lusa; Franco Mangiarotti ; Attilio Milanese; Rosario Principe, Juan Carlos Perez ; Herve Prin ; Dariusz Pulikowski; Eelis Tapani Takala; Emmanuele Ravaioli ; Ezio Todesco are with CERN, CH-1211 Geneva 23, Switzerland (e-mail: ezio.todesco@cern.ch).

P. Ambrosio, G. Apollinari and S. Feher are with Fermi National Accelerator Laboratory (FNAL), Batavia, IL 60510 USA.
P. Ferracin is with Lawrence Berkeley National Laboratory (LBNL), Berkley, CA 94720 USA.
Color versions of one or more of the figures in this paper are available online at http://ieeexplore.ieee.org.
Digital Object Identifier

of mechanical stress on the quench performance, the MQXFS6 short model quadrupole for the High Luminosity Upgrade was tested under an azimuthal pre-load ranging from 50 % to 100 % of the electromagnetic forces at nominal current [12,13], keeping the same axial preload to fully compensate electromagnetic forces in this direction. This paper presents the quench performance and describes the mechanical behavior of the magnet under the different stress conditions. The magnetic measurements of the first allowed multipole are also used as a diagnosis tool to see if any evidence of coil unloading can be seen from the field harmonics. The paper first gives a short introduction on the magnet design (Section II) and then deals with the experimental results of the test (Section III).

## II. MAGNET DESIGN AND PARAMETERS

The MQXF quadrupole for the High Luminosity Upgrade for the LHC has a 150 mm wide aperture and 11.4 T peak field in the conductor at the nominal gradient of 132.6 T/m, required for operating the LHC at 7 TeV. Operation at 7.5 TeV (called ultimate performance) is also considered to prove the operational margin of the magnet, that corresponds to a gradient of 142.1 T/m, a current of 17.5 kA and a peak field of 12.2 T. The magnet design relies on $Nb_3Sn$ conductor, and is based on the concepts and technologies developed in 2003-2013 LHC Accelerator Research and Development Program (LARP) [14]. In the following subsections we will outline the main features of the design (see Fig. 2), and we refer to [5,15] for a more complete description.

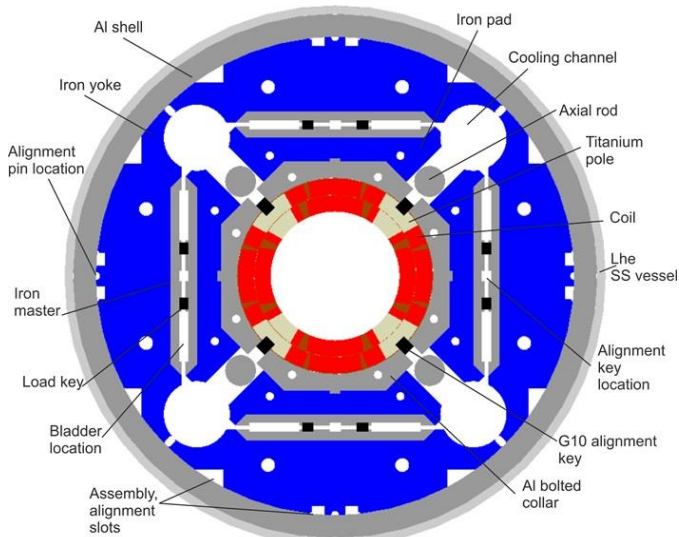

Fig. 2. Cross-section of the $Nb_3Sn$ low-β quadrupole magnet MQXF (including the LHe SS vessel, not present in the short models)

### A. Conductor

MQXF coils are made with a Rutherford-type cable composed of 40 strands of 0.85 mm diameter. The cable incorporates a 12-mm-wide, 25-μm-thick stainless-steel core to reduce inter-strand coupling currents. RRP 108/127 strands from OST-Bruker will be used for all series magnets. In the initial phase of the project, CERN also supported an effort to develop a conductor, namely the Powder-in-tube (PIT) strand by Bruker, with 192 subelements (see Fig. 3). A variant of the PIT conductor has been also developed in collaboration with CERN, introducing an additional Nb barrier around the whole bundle of filaments that allowed drastically reducing the effect of mechanical deformation and of the heat treatment cycle on the residual resistivity ratio (RRR) of the stabilizing wire copper [16]. Table I summarizes the main strand and cable parameters for the coils assembled in MQXFS6b/c/d, two based on PIT 192 strand and two made with the PIT 192 with bundle strand. Note that the measured strand critical current is ~5% lower than specification at 15 T (the value at 12 T is also given for reference).

All the coils were produced using the 2$^{nd}$ generation cable with a keystone angle of 0.4° [17]. The cable width and midthickness before reaction is 18.15 mm and 1.522 mm respectively. The cable is insulated with braided S2 glass, with a target thickness at 5 MPa of 0.145±0.005 mm [17].

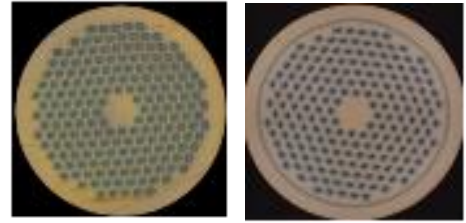

Fig. 3. Superconducting strands used for the MQXFS6b/c/d magnet. PIT 192 (left), PIT 192 with bundle barrier (right).

TABLE I: STRAND AND CABLE PARAMETERS

|  | Spec. | Coil 203 | Coil 204 | Coil 210 | Coil 212 |
|---|---|---|---|---|---|
| Strand layout |  | PIT 192 | PIT 192 | PIT 192bundle | PIT 192bundle |
| Strand non-Cu Jc (15 T, 4.2 K)$^2$, A/mm$^1$ | 1260 | 1238 | 1205 | 1216 | 1210 |
| Expected cable RRR | >100 | 108 | 105 | 117 | 120 |
| Strand non-Cu Jc (12 T, 4.2 K)$^2$, A/mm$^1$ | 2295 | 2279 | 2276 | 2147 | 2166 |

1 Measured in extracted samples reacted with the coil. Specification values include 5 % of the strand current density degradation due to cabling.

### B. Magnetic and coil design

MQXF coils produce a 132.6 T/m nominal gradient in a 150 mm aperture. Coil cross section is optimized for stress distribution among layers and field quality. Each coil is a double layer wound around a Ti alloy pole (Ti6Al4V), with 22 turns in the inner layer and 24 turns in the outer layer. Coil fabrication is based on the wind-and-react technology where the superconducting phase is formed after winding and during coil heat treatment. To accommodate the dimensional changes of the conductor during heat treatment, the cavity of the curing, reaction and impregnation tooling is designed accounting for 4.5 % cable expansion in thickness and 1.2 % in width [17]. The main magnetic parameters are summarized in Table II.

TABLE II.
MAGNET PARAMETERS OF SHORT MODELS

|  | Unit |  |
| --- | --- | --- |
| Operational temperature $T_{op}$ | K | 1.9 |
| Nominal/Ultimate gradient $G_{nom}/G_{ult}$ | T/m | 132.6/142.1 |
| Nominal/Ultimate current $I_{nom}/I_{ult}$ | kA | 16.23/17.50 |
| Nominal/Ultimate conductor peak field $B_{p,nom}/B_{p,ult}$ | T | 11.3/12.2 |
| Short sample current at 1.9 K | kA | 20.800 |
| Loadline fraction nominal/ultimate |  | 0.78/0.84 |
| Magnetic length | m | 1.196 |
| Differential inductance at $I_{nom}$ | mH/m | 8.21 |
| Stored energy at $I_{nom}$ | MJ/m | 1.17 |
| $F_x/F_y$ per octant at $I_{nom}$ | MN/m | 2.47/-3.48 |
| $F_z$ (whole magnet) at $I_{nom}$ | MN | 1.17 |

In Fig. 4 we show a magnification of the magnet load line around the short sample conditions at 1.9 K and 4.5 K, plotting the specified and the measured critical current surfaces. Therefore at nominal gradient/current, MQXFS6b is operating at 16.23/20.5 = 79% of the loadline at 1.9 K, and 85% at ultimate gradient/current, see Table III. In the following sections, we will express the magnet performance in terms of the fraction of the achieved current with respect to lowest short sample relative to the measured cables, i.e. (14.0 T, 20.5 kA) at 1.9 K and (13.0 T, 18.8 kA) at 4.5 K: this is a more relevant quantity for the community rather than the HL-LHC targets of nominal/ultimate current.

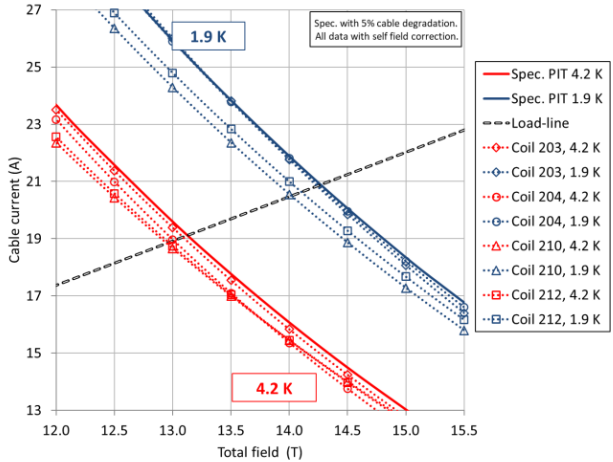

Fig. 4. Magnification of the cable critical current versus total magnetic field including self-field correction in the short sample region: values from strand specification, including 5 % of cabling degradation, fit curve of measurements performed on extracted strand data from MQXFS6b/c/d coils, and magnet load-line.

TABLE III: SHORT SAMPLE CURRENT FOR COILS USED IN MQXFS6B/C/D

| Short sample current (kA) | 1.9 K | 4.5 K |
| --- | --- | --- |
| Coil 203 | 20.840 | 19.060 |
| Coil 204 | 20.860 | 18.940 |
| Coil 210 | 20.507 | 18.819 |
| Coil 212 | 20.641 | 18.871 |

## C. Mechanical design and instrumentation

MQXF relies on a system of water-pressurized bladders and keys to apply a partial pre-stress to the coils and to pre-tension the aluminum shell (Al-7075) during loading at room temperature [18]. The azimuthal preload increase in the coils due to the differential thermal contraction of the structure components during the cool-down is 30 MPa, i.e. of the same order of magnitude of the preload overshoot required to insert the keys at room temperature. The baseline for magnet preload is not to exceed 110 MPa during the assembly at room temperature, aiming at an average of ~80 MPa after assembly, and 110 MPa at 1.9 K [19]. This target guarantees preload up to the nominal current of 16.23 kA producing the gradient of 132.6 T/m required for operating the LHC at 7 TeV. The mechanical design and analysis of the structure is described in detail in [5,15]. The coil titanium pole (see Fig. 2) is equipped with azimuthal and axial strain gauges to have an indication of the stress conditions in the coil. Strain gauges are mounted also on external surface of the Al shell, giving an additional measure which allows cross-checking the Ti pole gauges values. Both gauges are used to control the amount of preload applied to the coils during assembly, and to measure the strain state of the magnet components during cool-down and powering.

The axial preload is given via aluminium rods and end plates; the specification is to have a preload to compensate 100% of the axial forces induced during powering at nominal current (~1.2 MN, see Table II). Strain gauges on the axial rods allow to measure the tension state during the assembly, cool-down and powering.

## III. MQXFS6 ASSEMBLY AND POWERING TESTS

MQXFS6 is the fifth of the short model magnets, used to validate the design and performance in the first phase of the project [5,10,13]. MQXFS6 was assembled and tested at CERN, and was the second one assembling coils manufactured with the PIT conductor. During the first power test, the magnet reached nominal current of 16.23 kA at 1.9 K, corresponding to 79% of short sample current, but some detraining phenomena was observed, with maximum current oscillating between 16.0 and 16.5 kA, and all quenches were in the same coil 208. One of the possible causes of the performance limitation was the Residual Resistivity Ratio (RRR), well below the 100 specification both in coil 208 and 209: for this reason, the two coils were replaced by coils 203 and 204, previously tested in the MQXFS5 magnet that reached the 17.5 kA ultimate current, corresponding to 85% of short sample current [21]. This new assembly was called MQXFS6b. MQXFS6 was assembled with nominal preload both azimuthally and axially.

### A. Mechanical behavior during assembly and cool-down

The second assembly MQXFS6b aimed at evaluating the performance after the coil replacement and was assembled following the nominal target of ~80 MPa pole azimuthal stress after loading, and ~110 MPa at 1.9 K (i.e. preload at 80% of short sample). The next iteration (MQXFS6c) was to assemble the magnet with the minimum preload, which corresponds to a coil pole compression of ~30 MPa at room temperature after loading, and ~60 MPa at room temperature: this results in an azimuthal preload at 1.9 K corresponding to 60% of short sample. Lower preload was not possible due to the risk of damaging the magnet during handling operations. In Fig. 5 we show the variation of azimuthal stress in the shell and in the pole during loading, for the different loading key steps. In Fig. 5, the data are compared to the mechanical finite element model, finding the same slope. Note that the magnet was assembled without G10 pole keys, used to guarantee the alignment, to remove one element that could have added a complexity not needed for this experiment. After the power test of MQXFS6c, the magnet was assembled with the load keys used for MQXFS6b to verify the reproducibility of performance (MQXFS6d). The values of the measured stress in the coil in the different assemblies are given in Table IV. The typical variation of preload from coil to coil is of the order of ±10 MPa.

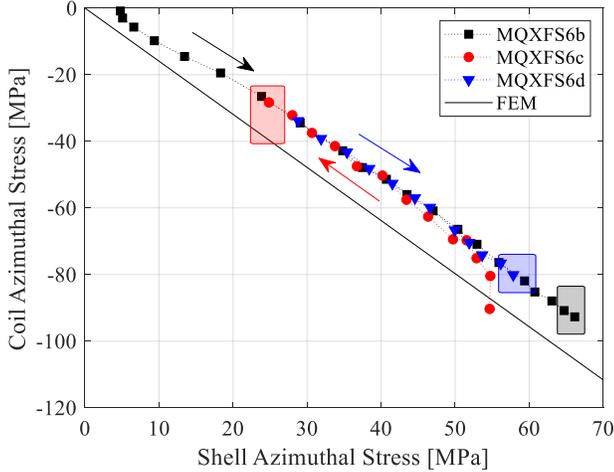

Fig. 5. Measured shell azimuthal stress versus measured coil azimuthal stress during coil loading; squares indicates the final status after assembly of the four coils; FEM is the finite element model result.

TABLE IV. KEY SIZE AND MEASURED COIL AZIMUTHAL STRESS

|  | Unit | S6b | S6c | S6d |
|---|---|---|---|---|
| Loading key size | mm | 13.85 | 13.6 | 13.85 |
| Coil 203 stress RT | MPa | -81 | -41 | -74 |
| Coil 204 stress RT | MPa | -90 | -23 | -85 |
| Coil 210 stress RT | MPa | -100 | -26 | -84 |
| Coil 212 stress RT | MPa | -102 | -23 | -77 |
| Average coil stress, RT | MPa | -93 | -28 | -80 |
| Average shell stress, RT | MPa | 66 | 25 | 58 |
| Coil 203 stress 1.9 K | MPa | -98 | -64 | -114 |
| Coil 204 stress 1.9 K | MPa | -118 | -70 | -115 |
| Coil 210 stress 1.9 K | MPa | -119 | -66 | -109 |
| Coil 212 stress 1.9 K | MPa | -119 | -65 | -110 |
| Average coil stress, 1.9 K | MPa | -113 | -66 | -112 |
| Average shell stress, 1.9 K | MPa | 128 | 82 | 133 |
| Axial force, 1.9 K | MN | 1.13 | 1.09 | 1.11 |

## B. Mechanical behavior during powering

As the current increases, the electromagnetic forces gradually pull the coil away from the pole. In Fig. 6 we plot the evolution of the measured difference in the azimuthal pole stress during powering, measured with the strain gauges on the titanium pole. Initially, the stress linearly decreases with the applied forces (proportional to the square of the current). After a certain level, the linear behavior is lost: this is considered an indication of unloading of the coil from the pole. The data are consistent with the loading strategy, showing beginning of unloading at ~80% of short sample for MQXFS6b and MQXFS6d, and beginning of unloading at ~60% of short sample in MQXFS6c. Note that (i) the spread in the measured values between the coils is order of ±10 MPa and (ii) the slope of the change of stress with the square of the current is identical for all the assemblies.

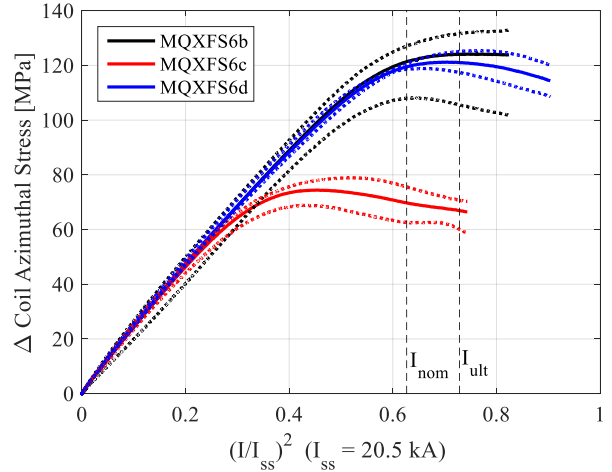

Fig. 6. Unloading during powering. Average (continuous lines) and variation across the four coils (dashed lines).

## C. Quench performance

As shown in Fig. 7, MQXFS6b reached 79% of short sample (nominal current) without quench at 1.9 K and 4.5 K; note all coils were already reaching these current, namely coils 210 and 212 in MQXFS6, and coils 203 and 204 in MQXFS5. After a thermal cycle, the magnet trained up to 19.1 kA (93% of short sample) at 1.9 K, corresponding to 13.2 T peak field. At 4.5 K, the magnet was trained up to 18.1 kA (96% of the short sample). A third thermal cycle was done to verify the training memory: the results were excellent, and the magnet reached 93% of short sample at 1.9 K after one quench.

After the decrease of preload, MQXFS6c had a first quench at 82% of short sample, whereas the first quench in MQXFS6b was at 87% of short sample: this indicates that pre-load at 60% of short sample (as used in MQXFS6c) increases training for currents larger than 80% of short sample. After 6 quenches, the magnet reached 85% of short sample, demonstrating that this target of the HL-LHC project can be reached with half of the nominal preload. After 20 quenches, the magnet reached 89% of short sample at 1.9 K. This suggests that the operational region of 90-95% of short sample is not accessible or requires much longer training when the lower preload of MQXF6c is used.

At 4.5 K, the performance is very close to MQXFS6b. The magnet showed excellent memory after thermal cycle despite the lower preload, showing that the training was not finished, and reaching after thermal cycle 90% of the short sample limit at 1.9 K.

After an increase of the azimuthal pre-load, back to the level of MQXFS6b, MQXFS6d went back with one quench to the record current in MQXF magnets of 19.57 kA, i.e., 95 % of the short sample limit at 1.9 K and 13.4 T peak field.

A few notes about quench location: 70% of the quenches started in coils 203 and 204 (35% each coil), whereas only 30% of the quenches started in coils 210 and 212 (approx. 15% each coil). Most of the quenches started in the straight section of the inner layer pole turn, with only 3 quenches starting in the mid-plane block of the inner layer, 1 quench in the outer layer and 3 quenches in the transition from the inner to the outer layer. The level of pre-stress does not have an impact on the quench location.

### D. Magnetic Analysis

Due to the lower pre-load in MQXFS6c, in the presence of electromagnetic forces the pole turn is expected to detach more than in MQXFS6b and move towards the mid-plane with a wider amplitude. This is inducing a variation of $b_6$ that can be measured using rotating coils. We recall that the field quality in a quadrupole is described on terms of multipoles according to

$$B_y + iB_x = B_2 \cdot 10^{-4} \sum (b_n + ia_n) \left(\frac{x+iy}{R_{ref}}\right)^{n-1} \quad (1)$$

where $B_x$ and $B_y$ are the field components in Cartesian coordinates, $B_2$ is the reference field, and $b_n$ and $a_n$ are the normalized harmonics coefficients at the references radius $R_{ref}$ (50 mm for MQXF). A simple analytical estimate proves that this effect is well above the sensitivity of a magnetic measurement (order of 0.1 units of $b_6$). Assuming that the coil is a sector coil with an inner radius $r$ and coil width $w$, the impact of a variation of the coil pole angle for a sector quadrupole pole is

$$\Delta b_6 = 10^4 \frac{2R_{ref}^4}{3\sqrt{3}} \left(\frac{1}{r^4} - \frac{1}{(r+w)^4}\right) \frac{\Delta\theta_0}{\ln(1+w/r)} \quad (2)$$

For MQXF, the aperture radius $r$ is 75 mm and the coil width $w$ is 36 mm: therefore, 0.1 mm displacement on the pole gives 1.1 mrad variation of the pole angle and a variation of $b_6$ of 1.65 units. Since the difference on coil pre-stress at 1.9 K in MQXFS6b and MQXFS6c is ~45 MPa, assuming a coil elastic modulus of ~20 GPa and an infinitely rigid structure, the angle of the pole $\theta_0$ shall vary of 1.2 mrad, corresponding to 0.110 mm pole displacement and 1.8 units of $b_6$, i.e. well above the sensitivity of the magnetic measurement system.

To have a precise estimate of the coil displacement during powering, the results of the 2-D finite element model in ANSYS were exported to the 2-D magnetic model implemented in ROXIE, to assess the effects of coil deformation on field quality. The displacement map corresponds to the state of the coil after cool-down and powering at different current levels. The computed displacements in ANSYS were applied to every strand of the magnetic model. The computation gives a difference in the azimuthal displacement during powering between the two assemblies of 0.03-0.04 mm, which is about one third of the analytical estimation assuming an infinitely rigid structure. In Fig. 8 we show the difference in the measured $b_6$ versus current between MQXFS6b and MQXFS6c: the effect is clearly visible, with the expected sign and order of magnitude. The comparison to the model shows a measured variation of 0.37 units versus an expected one of 0.47 units.

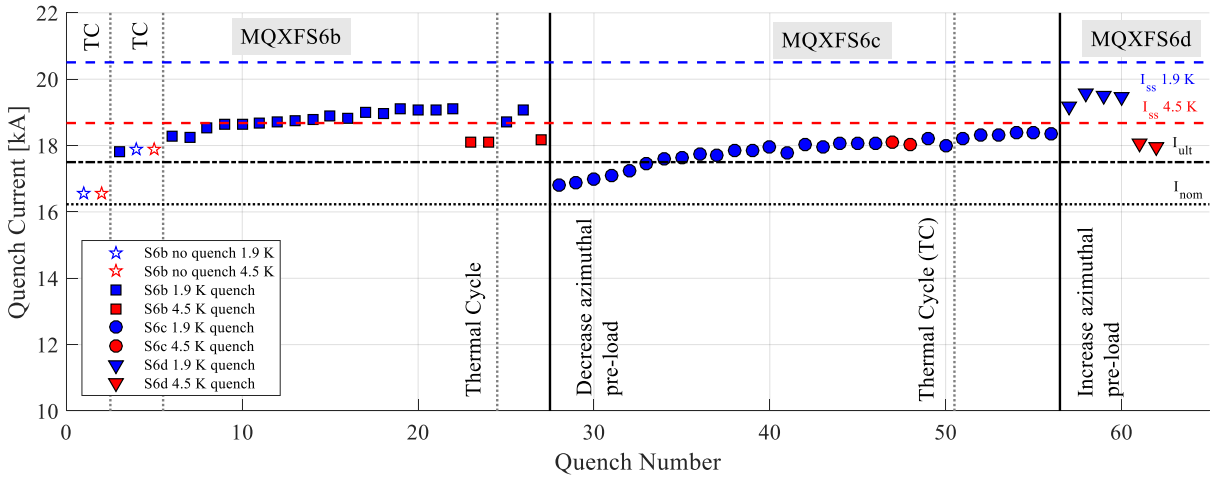

Fig. 7. Quench current of MQXFS6b-d. Ramps are at 20 A/s unless indicated. Data are compared to nominal, ultimate and short sample current (estimated from witness samples)

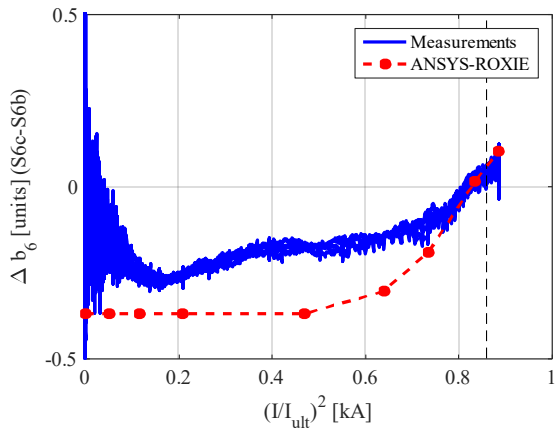

Fig. 8. Measured difference on $b_6$ in between MQXFS6b and MQXFS6c, due to wider detachment of the coil from the pole, as a function of the square of the current, normalized to the ultimate current (17.5 kA) compared to the ANSYS-ROXIE model.

## IV. CONCLUSIONS

In this paper we present an experimental verification of the influence of coil preload on magnet performance for the specific case of the $Nb_3Sn$ quadrupole short model developed in the framework of the HL-LHC project. The required level of preload for an accelerator magnet is debated in the community since many years. Even though the paradigm of the full preload is used as a guideline for most of the magnet designs, experience showed that the magnets can operate at currents sizably larger than the values corresponding to the pole unload.

MQXF magnet is based on a bladder and key structure that allows to vary the preload without a full disassembly, and a control of the level of preload in the range of ±10 MPa. In our experiment, we started with coils preloaded to compensate the electromagnetic forces arising at nominal current in HL-LHC (~80 MPa at room temperature, ~110 MPA at 1.9 K), corresponding to 80% of short sample (MQXFS6b). As expected, the magnet was able to operate to much higher currents, and reached 93% of short sample at 1.9 K. We then applied the preload required for 60% of short sample (~30 MPa at room temperature, ~60 MPa at 1.9 K): the magnet was still able to operate at 80% of short sample without retraining, but maximum current reached after a long training was 90% of short sample. The magnet was then warmed up and preloaded with the previous target, recovering 95% of short sample with one quench.

The experiment proves that (i) a preload to 60% of short sample allows operating the magnet at 85% of short sample without significant training and (ii) a preload at 80% of short sample allows to expand the operational field of the magnet from 90% to 95% of short sample current, and to reduce the training. The lower preload corresponds to a 0.040 mm larger detachment of the coil from the pole, as it is measured indirectly via the imperfections of the magnetic field, in agreement with finite element simulations.

ACKNOWLEDGMENTS

The authors wish to thank the technical staff for the construction and cold powering test of the magnet.